\documentclass{article}

\usepackage{fancyhdr}
\usepackage{lipsum}
\usepackage{longtable}
\usepackage{tcolorbox}  % For creating colored boxes

\usepackage{hyperref}

\hypersetup{
    colorlinks=true,
    linkcolor=blue, % choose a suitable color for hyperlinks
    urlcolor=blue,
    citecolor=blue
}

\usepackage{placeins}
\usepackage{float}
\usepackage{multirow}
\usepackage{graphicx} % For including images
\usepackage{amsmath} % Advanced math typesetting
\usepackage{tikz} % Drawing graphics
\usepackage{pgfplots} % Creating plots
\usepackage{booktabs} % Enhanced tables

\usepackage[T1]{fontenc}
\usepackage{mathptmx}

\usepackage{parskip}

\usepackage [autostyle, english = american]{csquotes}
\MakeOuterQuote{"}

\usetikzlibrary{patterns, intersections} % TikZ libraries for patterns and intersections
\usepgfplotslibrary{fillbetween} % PGFPlots library for filling between plots

\newcounter{myindex}

\pgfplotsset{compat=1.18} 
 
\title{Short note on the mapping of heritage sites impacted by the 2024 floods in Valencia, Spain}
\author{Josep Grau-Bové \thanks{Corresponding author: \href{mailto:josep.grau.bove@ucl.ac.uk}{josep.grau.bove@ucl.ac.uk}}, Richard Higham, Scott Orr, Pakhee Kumar \\ \small{Institute for Sustainable Heritage, University College London} }

\begin{document}

\date{}
\maketitle

\begin{abstract}
This short note presents preliminary findings on the impact of the October 2024 floods on cultural heritage sites in Valencia, Spain. Using publicly available data, we assess the extent of potential damage by overlaying flood maps with heritage site coordinates. We identify that 3.3\% of heritage sites in the region have been potentially impacted, with churches and shrines (81), outdoor religious iconography (78), and historic irrigation features (45) being the most heavily affected. Our analysis utilizes data from OpenStreetMap and listings from the Generalitat Valenciana, suggesting that while OpenStreetMap's crowd-sourced data can provide useful estimates of the proportion of impacted sites, it may not be suitable for a detailed damage assessment. By sharing this data openly, we aim to contribute to international efforts in preserving cultural heritage after the disaster and provide a foundation for future assessments of heritage site vulnerability to climate-related events.

\end{abstract}

\section{Introduction}

In early November 2024, devastating flash floods struck the Valencia region of eastern Spain, primarily affecting the southern outskirts of Valencia city and surrounding areas. The floods, which began on October 29th, were caused by intense rainfall that led to the overflow of the Magro and Turia river basins, as well as the Poyo riverbed. The disaster resulted in at least 219 fatalities, extensive damage to infrastructure, homes, and businesses.

We have written this brief note to present preliminary findings on the heritage impacts of the 2024 floods in Valencia. We are sharing this data openly, hoping it may save time to others and contribute, in some small way, to international efforts in solidarity with the affected communities. The data, and the files used for the analysis, are available here:

\begin{tcolorbox}[colframe=blue!60!black, colback=blue!10!white, coltitle=black, sharp corners]
    \textbf{Visit the project repository:} \\
    \url{https://github.com/jgraubove/danamaps}
\end{tcolorbox}

Rapid responses to heritage sites in crisis have increased in recent decades, supported by digital technologies like GIS, and often involving a component of crowdsourcing and volunteer-led efforts. The work of the Heritage Guard Network,a collaborative project between Wikimedia Sweden, Wikimedia Poland, Wikimedia Ukraine and Wikimedia Georgia, is a prominent and recent example \cite{heritage_guard_network}. Other initiatives worth noting are Kathmandu Cultural Emergency Crowdmap, an initiative led by ICCROM during the 2015 Nepal earthquake \cite{tandon2017post} and an initiative led by Wikimedia after the 2018 National Museum of Brazil fire \cite{Peschanski2018}. As the climate crisis intensifies, these initiatives will be more necessary. 

A scientific objective of this paper is to determine whether open, crowd-sourced data sources are useful for an initial appraisal of the potential damage to heritage.

\section{Methodology}

We used flood maps from the Copernicus Emergency Management Service, selecting the coverage from the first day of the event to capture the largest affected area \cite{CopernicusRapidMapping2024}.
    
 The heritage site coordinates were obtained from:
 
    \begin{itemize}
        \item \textbf{OpenStreetMap (OSM):} Crowd-sourced heritage locations and typologies from a global volunteer community, accessed via the OSM API \cite{OpenStreetMapAPI}.
        \item \textbf{Generalitat Valenciana Website:} Official listings of heritage sites as "Bens d'Interès Local" and "Bens d'Interès Cultural." Coordinates were extracted by web scraping, downloading the html code of the website where a siute map is embedded, and extracting the locations. It should bentoed that these are pending revision by the Generalitat, according to the website \cite{GeneralitatValenciana}.
    \end{itemize}

The obtained data contains very sparse metadata, limited to coordinates, name, and occasionally the address. 

The combined heritage coordinates were overlaid on the flood map using  using ArcGIS® software by Esri \cite{Ersrisoftware2024} and shown with a world topographic basemap\cite{basemap2024}. A site was flagged as impacted if located within 10 meters of the affected area.

\section{Results}

\subsection{Overall count of potentially affected heritage}

According to the most comprehensive list used in this assessment (the total number of sites listed by the Generalitat Valenciana), 3.3$\%$ of the heritage assets of the region are within the affected area (Table \ref{table:sum}). 

The Open Street Maps (OSM) data has proven to be substantially less populated than the official records. Nonetheless, it has provided an excellent estimate of the number of "Interes Cultural" sites affected. These sites can be considered, simplistically, to be of higher value (or higher listing status) than "Interes Local". 

If the purpose of a rapid asessment is to identify the proportion of sites affected in a region, sampling through OSM can be a good approach. This could be useful in cases where comparisons between regions are of interest. 

It is interesting to consider this fact in the light of sampling theory. To determine a proportion of  3$\%$ with an acceptable error of 1$\%$ and a confidence of  95$\%$ sampling theory indicates that the necessary sample size is approximately 1120. This gives a rough indication of the minimum number of OSM locations in a region that should be expected for a good estimation of the fraction affected. This simplified calculation assumes that the locations are homogeneously distributed.

\begin{table}[h!]
\centering
\begin{tabular}{|l|r|r|r|}
\hline
\textbf{Source} & \textbf{Total Listed} & \textbf{Within 10 m of Affected Area} & \textbf{Percentage} \\
\hline
Open Street Maps & 3544 & 51 & 1.44 \\
Generalitat Total & 12921 & 427 & 3.30 \\
Generalitat Interes Local & 11807 & 411 & 3.48 \\
Generalitat Interes Cultural & 1114 & 16 & 1.44 \\
\hline
\end{tabular}
\caption{Data Summary: affected sites according to different data sources}
\label{table:sum}
\end{table}

\subsection{Maps}

\begin{figure}[H]
    \centering
        \includegraphics[width=1\textwidth]{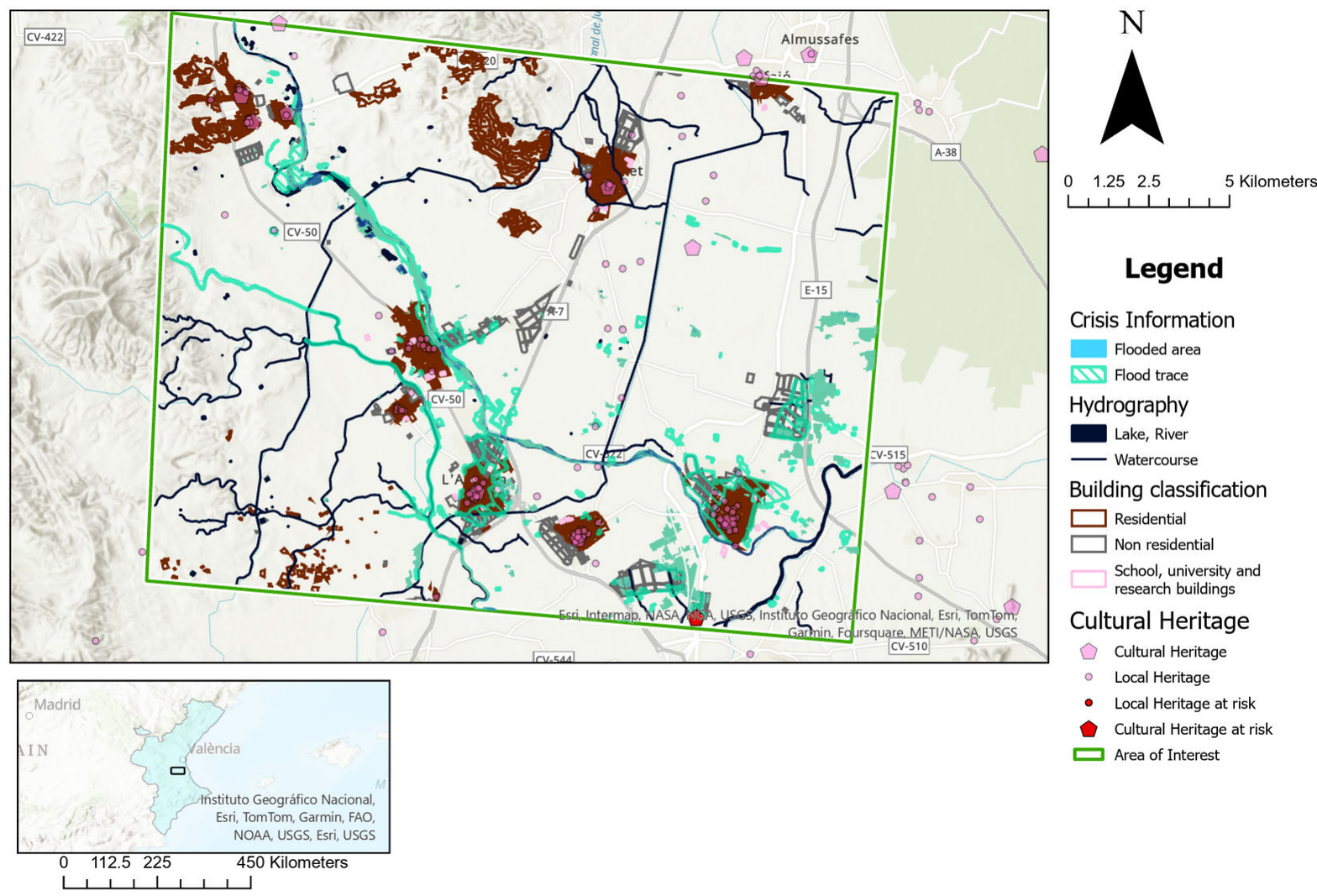} % Adjust width as needed
    \caption{Map of Algemies}
    \label{fig:example4}
\end{figure}

The maps below show the location of the potentialy impacted heritage sites in the regions of Valencia, Horta Sud and Algemies. Figure \ref{fig:example4} shows an example of a zoomed-in map. Others can be easily created if necessary to allow field inspection. 

\begin{figure}[H]
    \centering
    \includegraphics[width=1\textwidth]{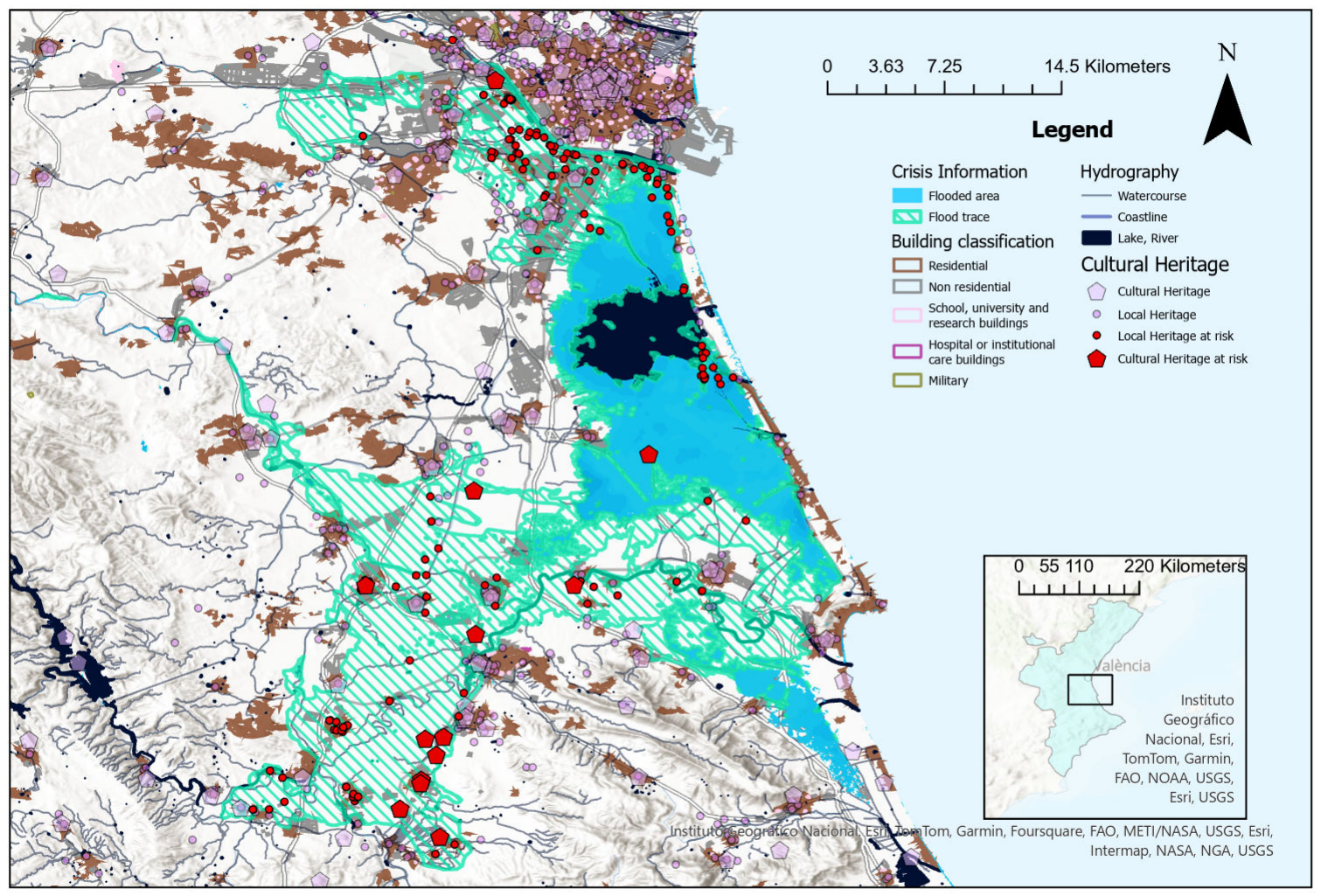} % Adjust width as needed
    \caption{Map of the region of Valencia }
    \label{fig:example1}
\end{figure}

\begin{figure}[H]
    \centering
        \includegraphics[width=1\textwidth]{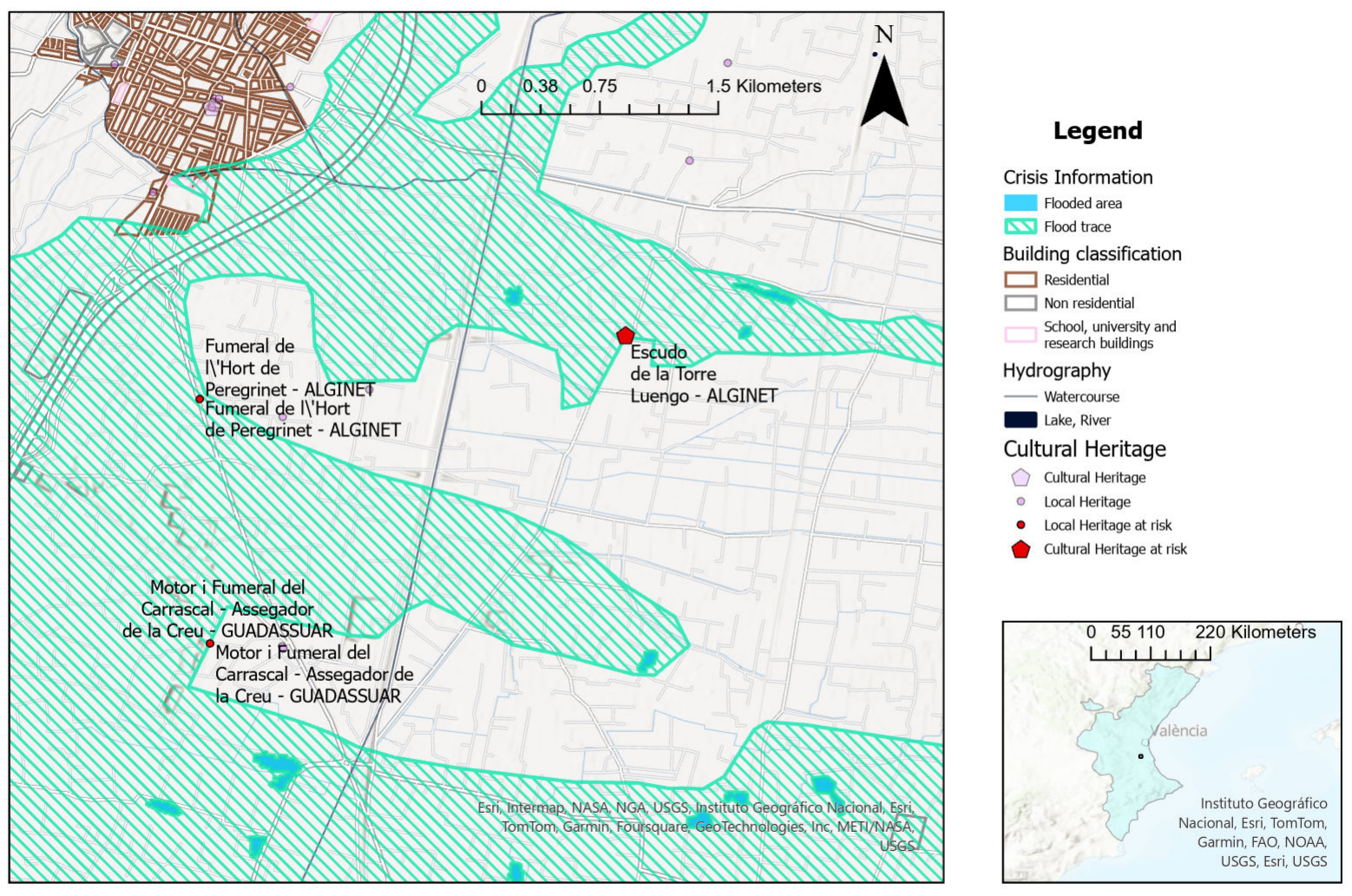} % Adjust width as needed
    \caption{Map of the region of Valencia with zoom to illustrate the ability of the GIS dataset to locate and label specific sites for inspection.}
    \label{fig:example2}
\end{figure}

\begin{figure}[H]
    \centering
        \includegraphics[width=1\textwidth]{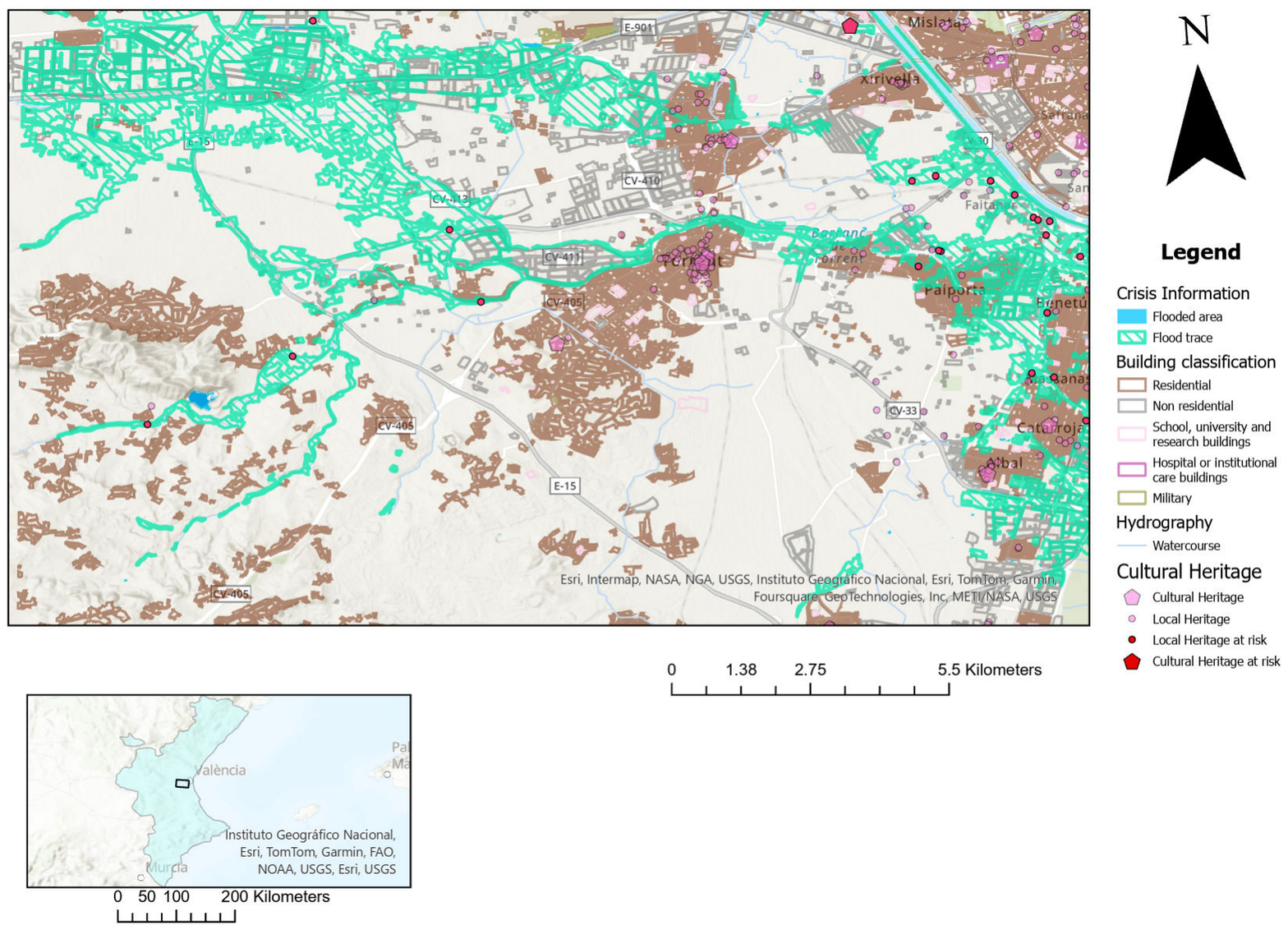} % Adjust width as needed
    \caption{Map of Horta Sud}
    \label{fig:example3}
\end{figure}

\subsection{Lists of potentially affected heritage}

While the dataset contains very sparse metadata, we used the names of the sites to provide a classification of the affected heritage typologies among the "Bens d'Interest Local" (Local Interest). This data is visible in Table \ref{table:long}. The "Bens d'Interes Cultural" (Cultural interest, higher listing) are less numerous and therefore are listed in full in Table \ref{table:cult}. 

Not surprisingly, there are numerous historic irrigation features (45), which are common in this area. Most concerning is the fact that the most commonly affected typology are Churches and Ermitas (small devotional buildings, usually translated as "shrine"), which total 81. This is followed by Religious Iconography (78), a broad category that includes outdoors religious imagery, for example made by ceramic tiles. Much of this iconography is placed on walls at a height of 1 or more meters, but this may not serve as a reliable assessment of vulnerability. 

\begin{table}[H]
\centering
\begin{tabular}{|l|l|}
\hline
\textbf{Name} & \textbf{Location} \\
\hline
Tramo Histórico de la Acequia de Mislata & Quart de Poblet \\
Cruz Cubierta de Alzira & Alzira \\
Iglesia Parroquial de San Andrés Apóstol & Alcúdia (L') \\
Escudo de la Torre Luengo & Alginet \\
Cruz Cubierta de Alzira & Alzira \\
Ermita de San Roque de Ternils & Carcaixent \\
Escudo de los Talens Albelda & Carcaixent \\
Iglesia Parroquial de San Bartolomé & Carcaixent \\
Escudo de los Almunia Esparza & Pobla Llarga (La) \\
Escudo de los Almunia Esparza &  \\
Escudo de los Brú & Pobla Llarga (La) \\
Casa del Lloc de Sinyent & Polinyà de Xúquer \\
Tramo Histórico de la Acequia de Mislata & Quart de Poblet \\
Recinto Amurallado de Berfull & Rafelguaraf \\
Torreón & Sant Joanet \\
Creu de la Llonga & Sueca \\
\hline
\end{tabular}
\caption{List of Locations potentially impacted in the category "Bens d'Interes Cultural"}
\label{table:cult}
\end{table}

\begin{longtable}{|l|c|p{10cm}|}
\hline
\textbf{Type} & \textbf{Count} & \textbf{Examples} \\
\hline
\endfirsthead
\hline
\textbf{Type} & \textbf{Count} & \textbf{Examples} \\
\hline
\endhead
\hline
Religious iconography & 78 & Retaule Ceràmic de Sant Francesc de Paula, located at Pl. Major 34 - TORRENT. Retaule Ceràmic de la Mare de Déu dels Desemparats, located in ALBERIC. Retaule Ceràmic de Sant Josep, located at C/ Sant Josep 68 - TORRENT. \\
Industrial & 58 & Fumeral d'un Rajolar, located in Parking Forsán - MASSANASSA. Fumeral de Motor i Tancat del Pomero, located in VALÈNCIA. Chimenea de Nave Industrial Junto a Escuela de Capataces, located in CATARROJA. \\
Church & 52 & Iglesia Parroquial de San Lorenzo Mártir, located in ALBERIC. Iglesia Parroquial de Nuestra Señora de la Salud, located in XIRIVELLA. Iglesia Parroquial de San Roque, located in BENICULL DE XÚQUER. \\
Irrigation & 45 & Caño de Catarroja - Barranco de Chiva o Torrente, located in CATARROJA. Assut del Barranc de l'Horteta, located in TORRENT. Llengües de Paiporta o de Faitanar, located in VALÈNCIA. \\
Historic House & 43 & Alqueria de Ferrer, located in VALÈNCIA. Alquería de Rocatí, located in VALÈNCIA. Alqueria del Saboner, located in VALÈNCIA. \\
Ermita (Shrine) & 29 & Ermita de San Onofre Anacoreta, located in ALGEMESÍ. Ermita del Cristo de la Agonía, located in ALGEMESÍ. Ermita de San Sebastián, located in POLINYÀ DE XÚQUER. \\
Bridge & 16 & Puente de Hierro sobre el Júcar entre Albalat de la Ribera y Polinyà de Xúquer, located in ALBALAT DE LA RIBERA. Pont de la Partida de Velasco, located in ALZIRA. Puente de Jalance, located in REQUENA. \\
Via Crucis Cross & 12 & Via Crucis. Estació I, located in XIRIVELLA. Via Crucis. Estació VII, located in ALGEMESÍ. \\
Historic Path & 12 & Espacio de Protección Arqueológica Vía Augusta, located in CATARROJA. \\
Bunker & 10 & Refugio antiaéreo urbano en Picanya, located in PICANYA. Refugio antiaéreo urbano en Alberic, located in ALBERIC. Refugio antiaéreo urbano en Pobla Llarga, located in POBLA LLARGA. \\
Mill & 16 & Molí de Raga o de Benetússer (Fariner i Arrosser), located in BENETÚSSER. Molí de Pala, located in VALÈNCIA. Molí Arrosser de Salvador Benlloch, located in VALÈNCIA. \\
Convent & 8 & Antiguo Convento de Nuestra Señora de los Ángeles, located in ALBERIC. Colegio de Santo Domingo (Antiguo Convento), located in CASTELLÓ. Convento de Santa Bárbara, located in ALCÚDIA. \\
Archaeology & 8 & Espacio de Protección Arqueológica Villa romana Pou de la Sargueta, located in RIBA-ROJA DE TÚRIA. Espacio de Protección Arqueológica Alteret de Marinyet, located in GUADASSUAR. Àrea Etnològica 1 - Racó de l'Olla, located in VALÈNCIA. \\
Industrial Heritage & 5 & Fumeral d'una Antiga Fàbrica de Pintes, located at C/ Giménez y Costa - VALÈNCIA. Motor i Fumeral de Tortosa, located in GUADASSUAR. Fumeral de l'Hort de Peregrinet, located in ALGINET. \\
Historic Village & 4 & Núcleo primitivo de Castellar, located in VALÈNCIA. Núcleo primitivo de El Palmar, located in VALÈNCIA. \\
Burial ground & 4 & Cementeri Municipal, located in GUADASSUAR. Cementeri Municipal, located in SEDAVÍ. Cementeri Municipal, located in MASSANASSA. \\
Historic Hospital & 2 & Hospital de Carabineros en Villanueva de Castellón, located in CASTELLÓ. \\
Historic Airport & 2 & Aeródromo en Catadau, located in CATADAU. \\
Military Heritage & 2 & Polvorín en Alberic, located in ALBERIC. Refugio antiaéreo urbano en Picanya, located in PICANYA. \\
Pier & 2 & Embarcador i Varador de el Palmar, located in VALÈNCIA. \\
\hline
\caption{List of locations potentially impacted by typology, in the category "Bens d'Interes Local". }
\end{longtable}
\label{table:long}

\section{Next steps}

\begin{itemize}
    \item The authors are open to collaboration on similar research in support of local academics and authorities. 
    \item The natural next step is to confirm potential damage with actual observations on the field. Local volunteers and online volunteers (i.e. the Wikimedia community) are well placed to contribute to this task. The Institute for Sustainable Heritage can contribute expertise on crowd sourcing and automatic analysis of the collected data with machine learning.
    \item Any further comparisons with better (or worse) data can contribute to more refined knowledge on the right sources for rapid assessments

\end{itemize}

This short note will be updated if necessary with new versions in this repository.

\bibliographystyle{plain}

\begin{thebibliography}{1}

\bibitem{OpenStreetMapAPI}
OpenStreetMap Contributors.
\newblock Openstreetmap api, 2024.
\newblock Accessed: 2024-11-12.

\bibitem{Ersrisoftware2024}
Esri.
\newblock Arcgis pro (version 3.0) [software], 2024.
\newblock Accessed: 2024-11-13.

\bibitem{basemap2024}
Esri.
\newblock Topographic [basemap]scale not given. world topographic map., 2024.
\newblock Accessed: 2024-11-13.

\bibitem{heritage_guard_network}
Wikimedia Foundation.
\newblock Heritage guard network.
\newblock \url{https://meta.wikimedia.org/wiki/Heritage_Guard_Network}, n.d.
\newblock Accessed: 2024-11-12.

\bibitem{Peschanski2018}
João~Alexandre Peschanski.
\newblock After a catastrophic fire at the national museum of brazil, a drive to preserve what knowledge remains, 2018.
\newblock Accessed: 2024-11-13.

\bibitem{CopernicusRapidMapping2024}
Copernicus Emergency~Management Service.
\newblock Rapid mapping: Floods in valencia (emsr773), 2024.
\newblock Accessed: 2024-11-12.

\bibitem{tandon2017post}
Aparna Tandon.
\newblock Post-disaster damage assessment of cultural heritage: Are we prepared, 2017.

\bibitem{GeneralitatValenciana}
Generalitat Valenciana.
\newblock Inventario general del patrimonio cultural valenciano.
\newblock Accessed: 2024-11-12.

\end{thebibliography}

\end{document}